\documentstyle[11pt,newpasp,twoside,epsf]{article}
\def\edcomment#1{\iffalse\marginpar{\raggedright\sl#1\/}\else\relax\fi}
\reversemarginpar

\begin{document}
\title{Observational tests of the GAIA expected harvest 
on eclipsing binaries}
\author{Toma\v{z} Zwitter}
\affil{University of Ljubljana, Dept.\ of Physics, Jadranska 19,
1000 Ljubljana, Slovenia, tomaz.zwitter@fmf.uni-lj.si}

\begin{abstract}
GAIA observations of eclipsing binary stars will have a large 
impact on stellar astrophysics. Accurate parameters, including 
absolute masses and sizes will be derived for $\sim 10^4$ systems,
orders of magnitude more than what has ever been done from the ground. 
Observations of 18 real systems in the GAIA-like mode as well as 
with devoted ground-based campaigns are used 
to assess binary recognition techniques, orbital period 
determination, accuracy of derived fundamental parameters and the need to 
automate the whole reduction and interpretation process.
\end{abstract}

\section{Introduction}

GAIA observations of eclipsing binary stars will be of utmost importance 
to advances in stellar astrophysics. For no other class of objects
one could determine fundamental stellar parameters, i.e.\ absolute mass, 
size and surface temperature distribution with a 
comparable accuracy. Solutions of wide detached binaries can be 
used to accurately position them on the absolute H-R diagram. Identical 
age of both components places useful constraints on the theoretical 
isochrones for the given metallicity and rotational velocity which will also 
be derived from GAIA observations. Components in short period systems are 
closer and mutually disturbed, so their evolution is different from 
that of single stars. But accurate surface temperatures and sizes 
derived from binary solutions fix their luminosity and are useful to gauge 
their distance even for objects that are too far for the astrometric 
capabilities of the satellite (see Wythe \&\ Wilson 2002).

Availability of on-board spectroscopy is vital to the study of eclipsing 
binaries. Semi-major axis and stellar masses could not be determined in 
any other way, 
and additional information on metallicity and rotational velocity helps 
in physical interpretation. One might argue that this information could 
be obtained by ground-based follow-up observations. In our experience this 
is not feasible. In Asiago we launched an intensive campaign to 
spectroscopically observe eclipsing binaries discovered by Hipparcos 
(Munari et al.\ 2001 [hereafter M2001], Zwitter et al. 2003). After three years we barely 
finished the spectroscopic coverage of the first 18 systems. Hipparcos 
discovered nearly 1000 systems, GAIA will see hundred thousands. These objects 
are distributed over the whole sky, so fiber optic spectroscopy cannot 
reduce the required observing time significantly.

The strength of the GAIA mission is in the numbers. GAIA will observe 
$\sim 4 \times 10^5$ eclipsing binaries brighter than $V=15$,
$\sim 10^5$ of these will be double-lined systems (M2001). Even if the 
stellar parameters will be determined at 1\%\ accuracy only for 1\%\ 
of them this is still 25-times more than what has been obtained from all 
ground-based observations in the past (cf.\ Andersen 1991). Moreover most 
of the GAIA binaries will be of G-K spectral type (cf. Zwitter \&\ Henden 2003) 
where there exists only a small number of systems with accurate solutions. 

Astrophysical importance of eclipsing binaries is discussed in other 
contributions (Milone 2003, Wilson 2003, Van Hamme 2003). Here we 
focus on our experience obtained from real stars that were observed in 
the GAIA-like mode. We start with discussion of how an object is recognized 
to be a binary and a determination of its orbital period. Next we discuss the 
accuracy of derivation of its fundamental parameters and the possibility 
to detect intrinsic variability of stars in binaries. We close with 
some general remarks on the types of binaries that will be discovered. 
We stress that huge numbers of objects call for completely automated 
reduction and possibly even interpretation techniques. 
 
\section{Orbital period from multi-epoch observations}

A large fraction of binary stars with orbital periods over a month 
that are closer than 1~kpc will be discovered astrometrically. Systems 
with periods of up to 10~years will be recognized due to their non-linear 
proper motion and those with periods of over a century will be resolved 
(ESA SP-2000-4, Arenou 2003). Systems with orbital periods of less than 
a month will be mainly discovered by their photometric and spectroscopic 
variability. 

GAIA is unique because it will re-observe the same region of the sky 
many times over. The number of transits for the spectroscopic focal plane 
will be around 100 with extremes 
a factor 2 higher or lower. The transits are not distributed evenly in 
time (see Fig.\ 1). This should pose no problems in analysis of 
binary stars if the satellite rotation and precession periods are kept 
incommensurable. The sampling permits a good phase coverage of all 
orbital periods that are shorter than the mission lifetime. Also the duration  
of individual focal plane passages is just 100 seconds, so orbital motion 
smearing is negligible. 

Photometric variability does not need to be a consequence of the binarity of the 
source: pulsations, rotating and time-dependent stellar spots, as well as 
different types of semi-regular variables will be common among the G and K 
stars that will be the most frequent type of objects observed. The best 
way to recognize that the detected photometric variability is indeed due to binarity is 
by establishing its repeatability and light curve shape; so the orbital 
period needs to be determined. The same is true for spectroscopic 
observations. The exceptions are of course double-lined binaries where 
a quarter-phase spectrum with well separated lines immediately points 
to the binary nature of the source. 

Potentials of photometry and spectroscopy for determination of orbital 
period are different, with spectroscopy being always preferable. As an 
example let us examine a detached system GK~Dra (Fig.\ 2). It was 
discovered by the Hipparcos satellite. But a search for orbital period 
from the 124 Hipparcos observations proved unsuccessful. There are several 
possible periods with the most likely solution of 16.96-days (Fig.\ 2, top). 
This value, which is also quoted in the Hipparcos catalogue, is immediately shown 
to be wrong by only 35 spectroscopic observations in the GAIA spectral 
window obtained by our GAIA-like ground-based observing campaign (Zwitter 
et al.\ 2003; Fig.\ 2, middle). The advantage of spectroscopic monitoring 
is in the fact that radial 
velocities are constantly changing with orbital phase. So every point 
contributes to the orbital period search. In the case of photometry the 
light curve out of eclipses is flat, so determination of orbital period 
is based only on a couple of points within the eclipses. An extensive 
photometry can of course resolve this problem, but note how a periodogram 
from over 1300 photometric observations (Fig.\ 2, bottom) is still more ambiguous than 
the one obtained from only 35 spectroscopic observations in the GAIA spectral 
window. The strength of spectroscopy for the orbital 
period determination can also be seen from the light curves in Figure 3. 
 
\begin{figure}
\vspace*{23mm}
\plotfiddle{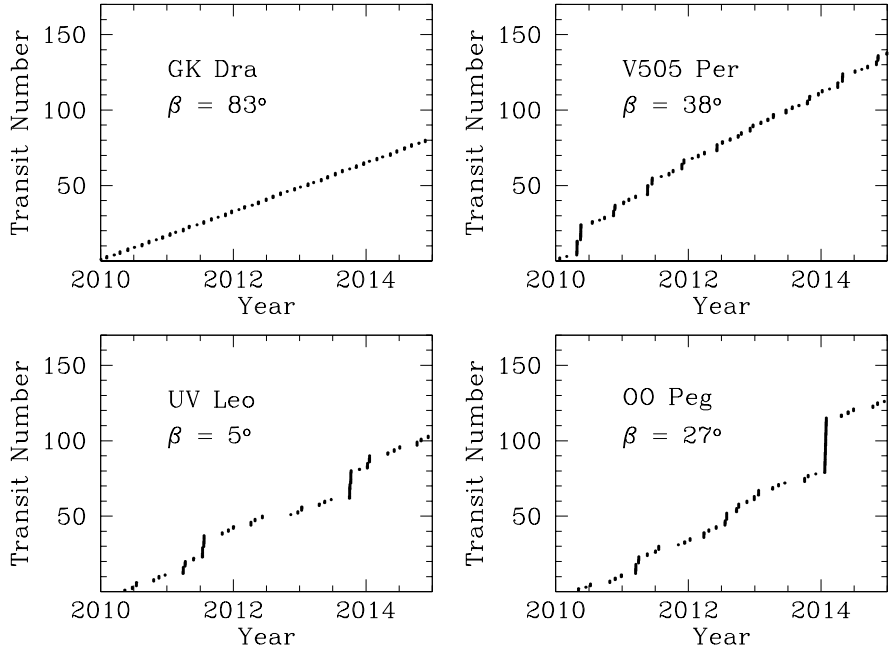}{6.0cm}{270}{46}{46}{-175}{270}
\caption{
Accumulation of transits over the spectroscopic focal plane 
during the 5-year mission lifetime for four examples of binary stars.
Dynamics of observations depends on the ecliptic latitude 
of the target~($\beta$). GK~Dra is located close to ecliptic pole, so 
the transits are almost periodic. Coverage of stars at other 
ecliptic latitudes is more patchy.
}
\end{figure}

\begin{figure}
\vspace*{33mm}
\plotfiddle{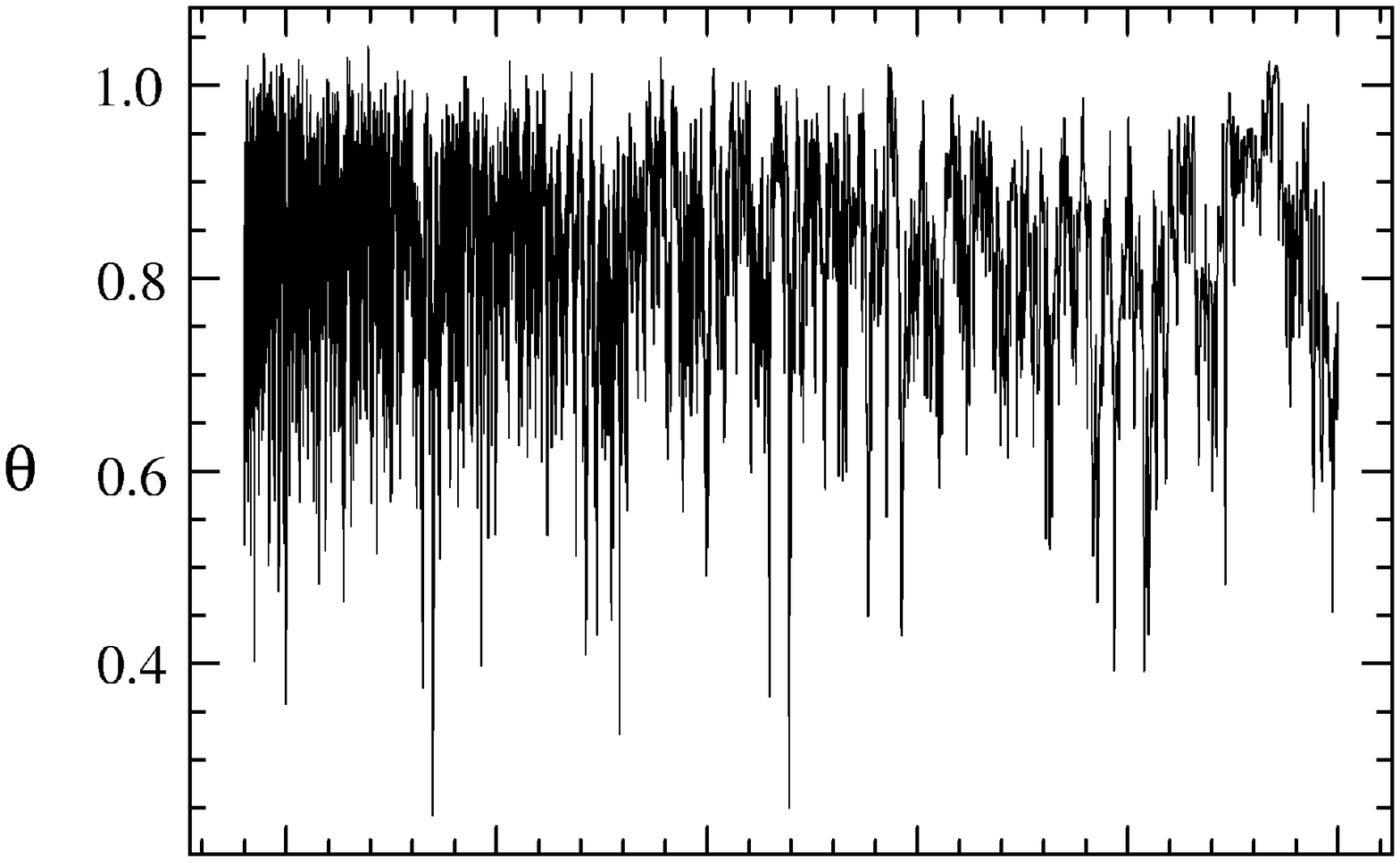}{3.52cm}{0}{49.5}{41.8}{-163}{48.7}
\plotfiddle{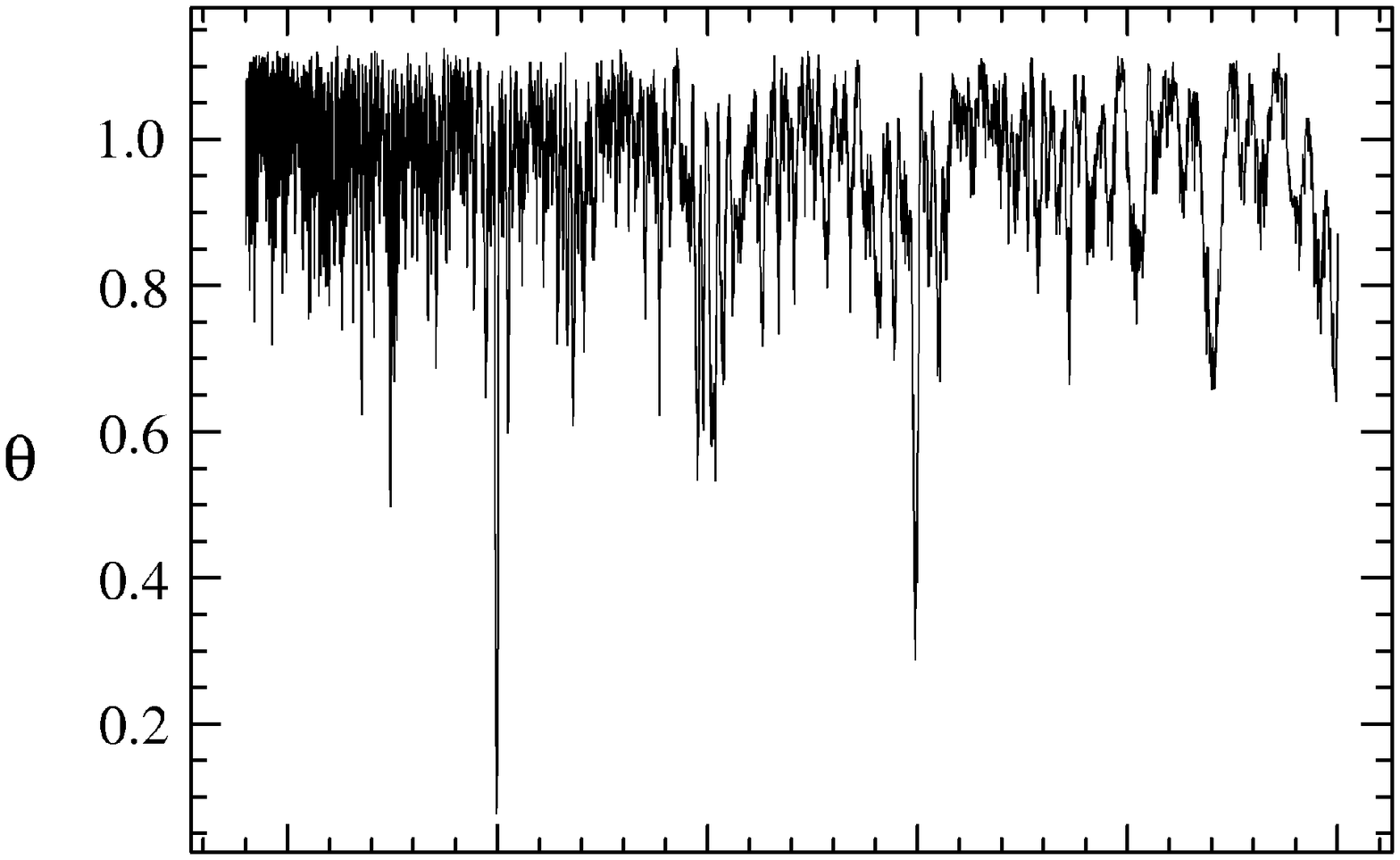}{3.52cm}{0}{49.5}{41.8}{-163}{10.2}
\plotfiddle{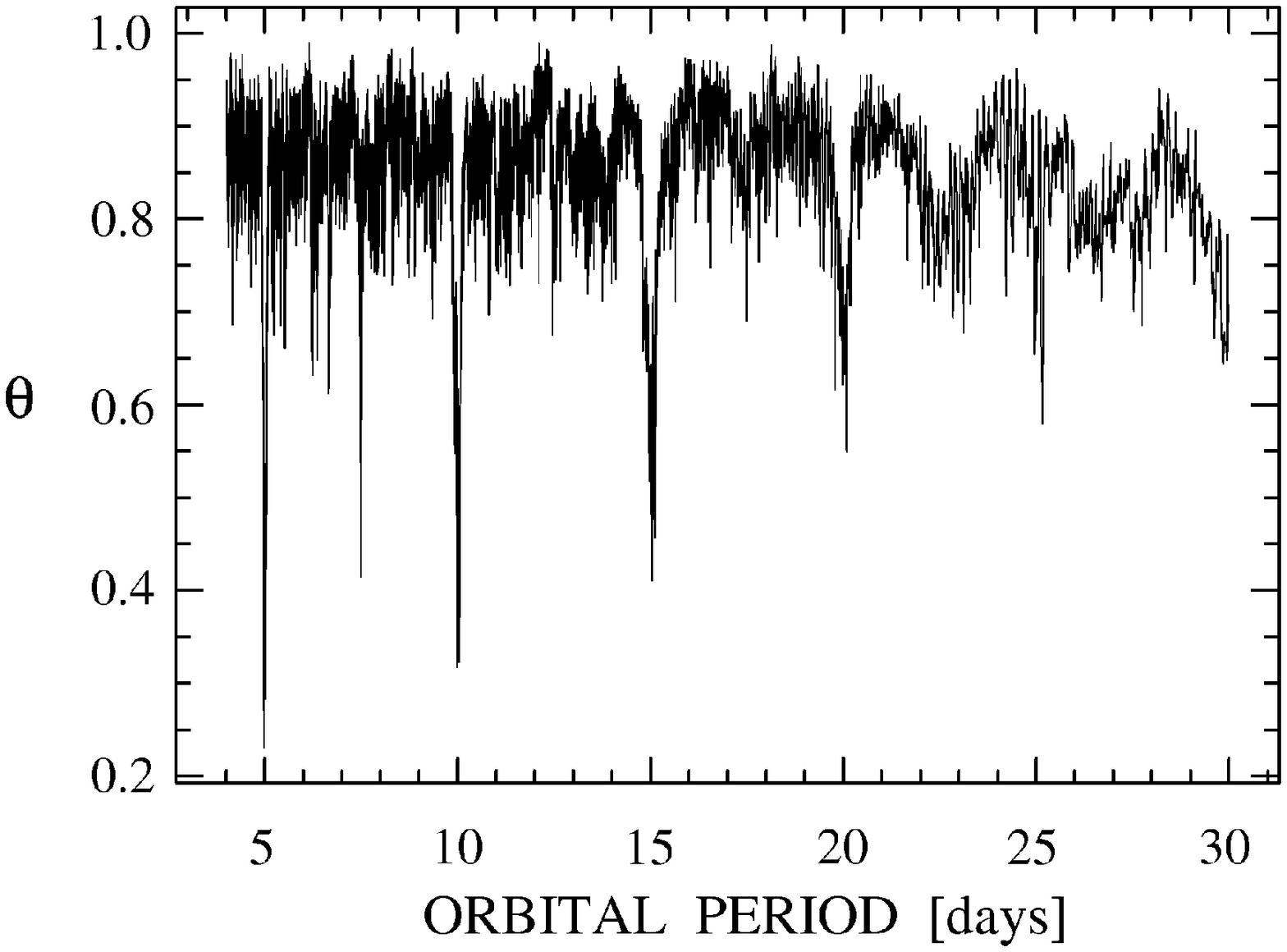}{3.96cm}{0}{54.2}{46.2}{-162}{-47}
\vspace*{17mm}
\caption{
Search for orbital period of GK Dra using a phase dispersion minimization 
method (Stellingwerf 1978) on three data sets. \\
{\it Top panel:} 124 Hipparcos observations in the H$_{\mathrm P}$-band. 
{\it Middle panel:} 35 spectroscopic measurements of radial velocity 
of the primary star. 
{\it Bottom panel:} 1323 ground-based V-band photometric observations
(Dallaporta et al. 2002). Hipparcos data favour the 16.96-day period, while
spectroscopy and dedicated photometry identify the correct value of
9.97-days.
}
\end{figure}

\begin{figure}
\plotone{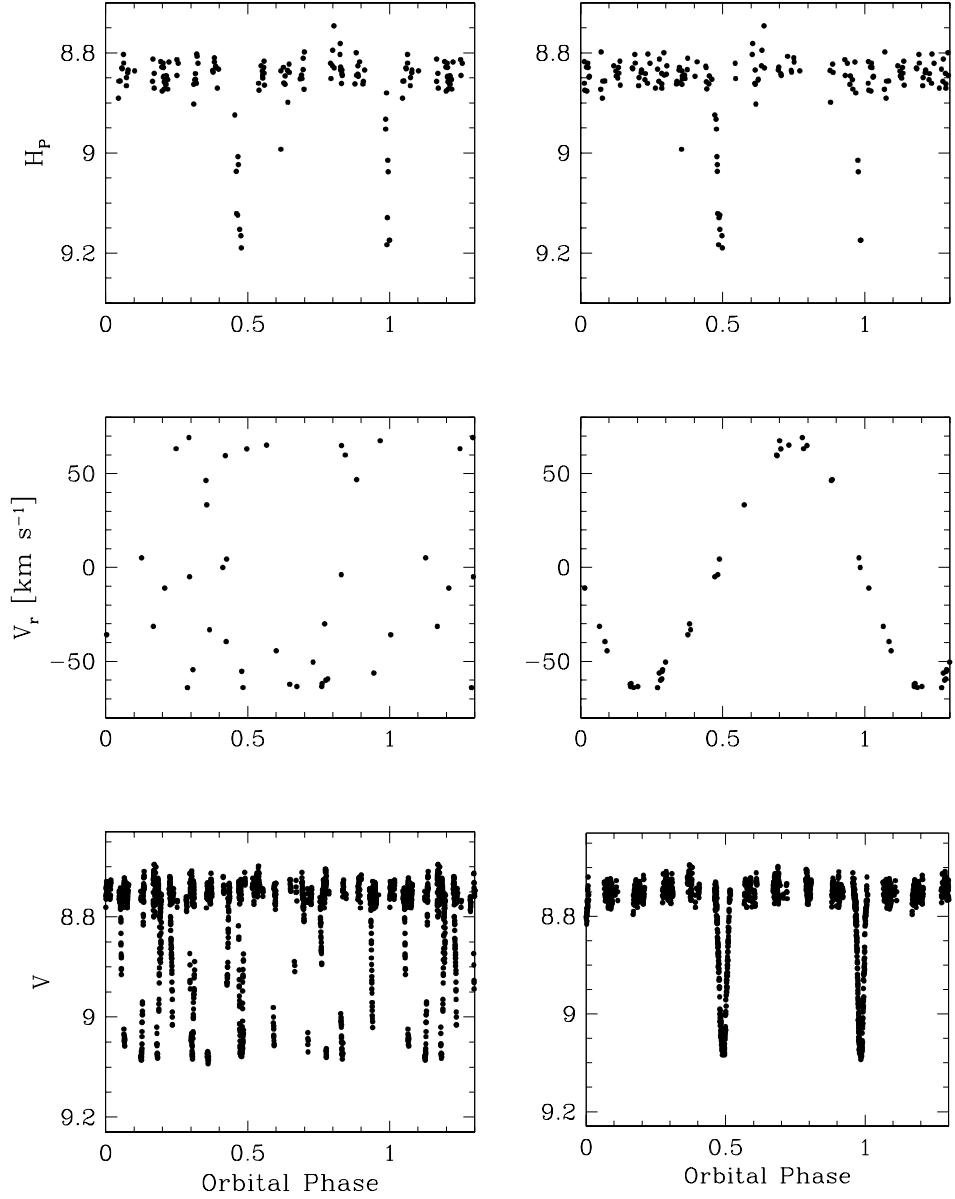}
\caption{
Orbital phase plots of GK Dra for orbital period of 16.96 days (left panels)
and 9.9742 days (right panels). 
{\it Top:} Hipparcos observations in the H$_{\mathrm P}$-band. 
{\it Middle:} spectroscopic measurements of radial velocity 
of the primary star. 
{\it Bottom:} ground-based V-band photometric observations.
}
\end{figure}

\begin{figure}
\vspace*{10mm}
\plotfiddle{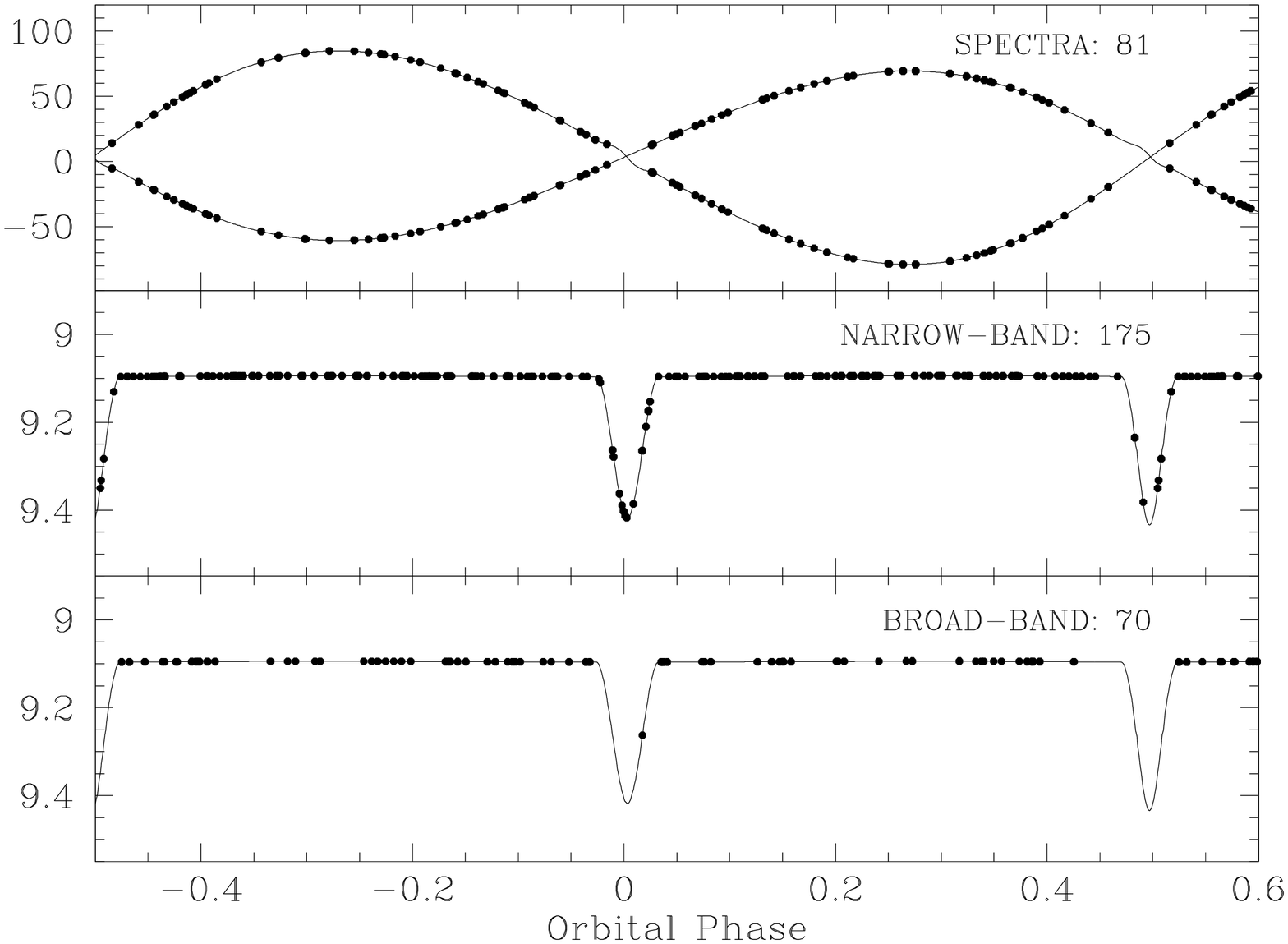}{4.5cm}{0}{48}{42}{-195}{-120}
\vspace*{16mm}
\caption{
Phase coverage of spectroscopic, narrow- and broad-band photometric 
observations for a typical eclipsing double-lined spectroscopic binary.
Data points mark individual passages of GK Draconis over spectroscopic, 
narrow- and broad-band filter focal planes between 1-Jan-2010 
and 31-Dec-2014.
Note a nice phase distribution of spectroscopic observations and a 
modest photometric coverage of eclipses.
}
\end{figure}

Photometric information obtained by the GAIA satellite will be far 
superior to that of Hipparcos. GAIA will reach much fainter magnitudes 
and observe in many broad and narrow photometric bands. But the number of 
epoch observations will be similar to that of Hipparcos. The coverage of 
GK~Dra to be obtained by GAIA is given in 
Figure 4. Note that only a single observation in broad band filters 
falls within eclipses. The eclipse coverage in intermediate passband 
filters is better. Broad- and narrow-band photometry 
can be used to measure orbital inclination 
as well as relative sizes and absolute temperatures of both stars. 
But the orbital period itself will be much easier to determine from 
spectroscopy. A total of 81 spectra are well distributed over orbital 
cycle, so the orbital period, semi-major axis, and both masses will 
be unambiguously measured from spectroscopic radial velocity measurements. 

\section{Accuracy of fundamental parameters}

As mentioned in the Introduction we are observing 18 Hipparcos 
binaries in the GAIA-like mode. This means we are trying determine their 
orbital solution and fundamental parameters using only Hipparcos ($H_P$,
$B_T$, $V_T$) photometry and ground based spectroscopy in the GAIA 
spectral range. Spectroscopic data are extracted from a single Echelle 
order observation with the  
1.8-m telescope of the Asiago observatory. It turns out that such an 
approach is realistic as the accuracy of the solution is limited by 
a rather small ($\sim 100$) number of Hipparcos photometric measurements 
resulting in a poor coverage of eclipses. This will be also the case 
with GAIA. The results we obtain should 
present a lower limit to the expected GAIA accuracy, as we are using 
only rather noisy Hipparcos photometry, while GAIA photometry will have 
excellent precision and a much larger number of photometric bands. Figure 5 
presents the light curve shapes of the overcontact binary V781 Tau 
obtained in different narrow and broad passbands. Note that the photometric 
accuracy for objects brighter than $V\sim 18$ will be better than 0.01 mag, 
so even rather subtle differences in the light curve shape and magnitude level 
for this 
binary with $T_1 - T_2 = 170 $~K will be easily discernible. 

\begin{figure}
\vspace*{14mm}
\plotfiddle{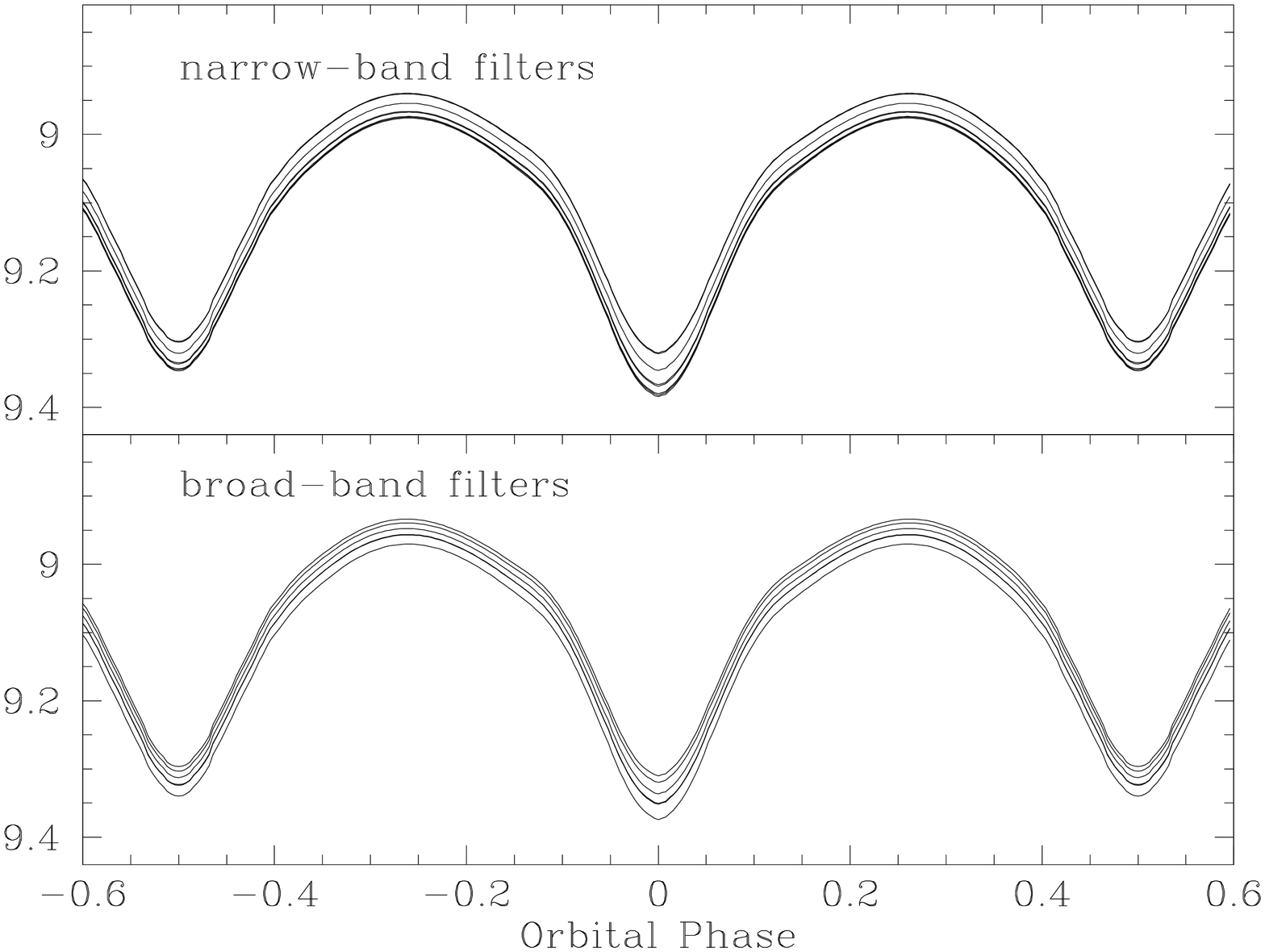}{5.0cm}{0}{45}{35}{-180}{-60}
\caption{
Simulated light curves of an overcontact binary V781 Tau for different 
narrow- (top) and broad-band (bottom) GAIA filters as defined in Munari (1999).
Each curve will be sampled with $\sim 100$ points with errorbars not 
exceeding 0.01 mag at V=18. 
}
\end{figure}

\begin{table}
\caption{Accuracy of fundamental parameters obtained from observations 
in the GAIA-like mode. Quoted errors are formal mean standard errors 
to the solution.}
\vspace*{3mm}
{\small
\begin{tabular}{l|llllll}
object                & V570 Per & OO Peg & V505 Per & V781 Tau & UV Leo & GK Dra \\ 
type		      & detached & detached&detached & overcontact&detached&detached\\
sp. type              & F5       & A2     & F5       & G0       & G0     & G0     \\ \hline
                      &          &        &          &          &        &        \\       
$a$                   & 0.7\%    & 0.5\%  & 0.5\%    & 0.2\%    & 1.2\%  & 0.5\%  \\
mass$_1$              & 2.3\%    & 1.7\%  & 1.5\%    & 1.8\%    & 7\%    & 3.5\%  \\
mass$_2$              & 2.5\%    & 1.8\%  & 1.6\%    & 1.8\%    & 6\%    & 3.3\%  \\
T$_1$                 & 150 K    & 150 K  & 40 K     & 50 K     & 100 K  & 100 K  \\
T$_2$                 & 180 K    & 180 K  & 60 K     & 30 K     & 100 K  & 100 K  \\
R$_1$                 & 10\%     & 4\%    & 1.4\%    & 0.4\%    & 2\%    & 1.5\%  \\
R$_2$                 & 25\%     & 4\%    & 3\%      & 0.3\%    & 2\%    & 1.7\%  \\
distance              & 6\%      & 6\%    & 7\%      & 1.5\%    & 20\%   & 10\%   \\ \hline
\end{tabular}
}
\end{table}

In Table 1 we quote accuracies of fundamental parameters for 6 systems
published so far (M2001, Zwitter et al.\ 2003). We note that relative errors in most parameters are 
2\%\ or lower. The exceptions are individual radii which are not well determined due to a 
scarce photometric coverage of eclipses. Solutions of UV Leo and GK Dra also have 
large uncertainties. This is due to their intrinsic variability (see below). V781 Tau 
is a binary which fills its Roche lobe up to the L2 point. Temperatures and sizes of such 
binaries can be accurately determined. So the distance can also be calculated with a remarkable 
accuracy.   

\section{Intrinsic variability}

Many stars of G and K spectral types are intrinsic variables. This is true also for binary 
members. With this goal in mind the observations in that GAIA-like mode that use only 
Hipparcos photometry with $\sim 100$ observations of each star were supplemented by devoted 
ground-based photometric campaigns (Dallaporta et al.\ 2000, 2002, 2002a, Mikuz et al.\ 2002, 
Frigo et al.\ 2002). 

UV Leo is a detached system with surface spots which cause a variation of system brightness by 
$\sim 0.04$~mag (Mikuz et al.\ 2002). The system also showed a sudden change in the orbital period 
in Feb.\ 1981, possibly due to a passage of a low-mass third body. Devoted photometry of GK Dra 
(Fig.\ 3, bottom right) shows unusually large scatter. It turns out that the differences 
between the binary solution and observations are not due to noise but point to 
an intrinsic variability of $\delta$-Sct type (Figure 6). 

A limited number of photometric and spectroscopic observations obtained by GAIA will make it difficult 
to study intrinsic variability of the binary components. Still a large number of photometric 
bands will easily point to temperature changes that do not repeat with orbital cycle and are so 
due to intrinsic variability of the binary components. Interesting cases could be picked for 
detailed follow-up observations. These include new interacting binaries (Cropper 2003).
 
\begin{figure}
\plotone{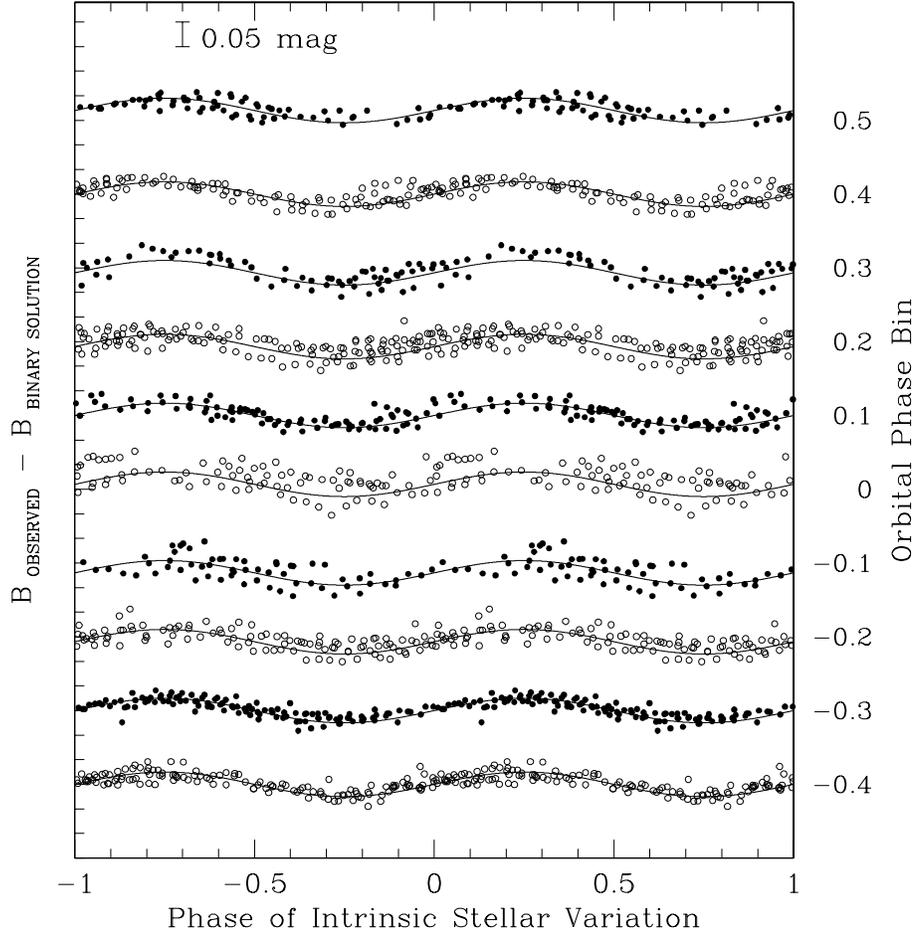}
\caption{
The difference between the observed B magnitudes of GK Dra (Dallaporta 
et al.\ 2002) and the ones generated from the binary system solution, 
folded on an 
intrinsic stellar variation period of $\sim 170$ minutes. Differences 
pertaining to different orbital phase bins ($P_{orb} = 9.97 $~days) 
are marked by different symbols and vertically offset for clarity. 
Note that a sinusoidal variation with a peak-to-peak amplitude of 
$\sim 0.05$ mag is present throughout the orbital cycle and is 
maintaining its phase. Intrinsic variability of binary components 
will be common among GAIA binaries.
}
\end{figure}

\section{Non-eclipsing and non-spectroscopic binaries}

So far we discussed eclipsing double-lined spectroscopic binaries. In majority of 
cases we will be less fortunate. The systems could be too faint to 
obtain any useful spectroscopy, non-eclipsing or single-lined. We discuss these in turn.

For systems fainter than $V=15$ spectroscopic 
radial velocities will be difficult to measure even in double-lined cases. These 
faint objects will far outnumber the bright spectroscopic binaries. Binarity will 
have to be established from eclipses or a reflection effect, both measured with a  
large number of photometric passbands but in a limited number of epochs. Some 
problems with the determination of orbital period of such systems have been discussed 
in Section 2. Here we only note that the binarity of these sources could be in general 
easily established due to a large photometric accuracy. In many cases the orbital 
ephemeris could also be derived, therefore permitting ground-based spectroscopic follow-up 
observations at quarter phases establishing absolute masses and dimensions of the 
binary components. The easiest to recognize will be systems close to contact, where 
much of the system information could be recovered from photometry alone.  

Eclipses will be rather uncommon, especially in the wide detached cases. But for a double-lined 
non-eclipsing spectroscopic binary the mass ratio can still be easily calculated. And 
in not too wide binaries the reflection effect can be measured from accurate photometry. 
This constrains the temperatures and inclination of the system. The inferred spectral 
classification can be finally checked against the properties of the spectra obtained 
close to quadratures where the spectral lines are well separated.    

Most binaries with the mass ratio below 0.3 will be single-lined, permitting to derive 
only a spectroscopic mass function. 

\section{Some remarks on reduction and interpretation procedures}

GAIA will discover huge numbers of spectroscopic and eclipsing binaries. The numbers 
are orders of magnitude larger than everything collected in the last century from the 
ground. In many cases the observations obtained by GAIA will be good enough to determine 
system parameters at 1-2\%\ accuracy level. They will have an immense impact on theories 
of stellar structure and evolution. 

Such a large data set requires an automation of all stages of reduction and interpretation. 
No-one could recognize photometric eclipses or winging radial velocity curves in hundred-thousands 
or even millions of systems by eye. But even interpretation and classification has to be 
completely automatic with only the most unusual cases to be marked for human inspection. 
Wythe \&\ Wilson (2001, 2002) successfully classified some photometric eclipsing binaries 
from the OGLE database with semi-automatic procedures. Prsa (2003) obtained some encouraging 
results for double-lined eclipsing binaries. Clearly 
development of reliable classification and analysis procedures is one of the major tasks 
facing the scientific community before the launch of GAIA.  
\\
{\bf Acknowledgement.} Generous allocation of observing time 
by Osservatorio astronomico di Padova and financial support from the Ministry of Education, 
Science and Sport of Slovenia are acknowledged.

\end{document}